 \documentclass[11pt,a4paper]{article}

\usepackage[T1]{fontenc}
\usepackage{color}
\usepackage{amssymb}
\usepackage{amsmath}
\usepackage{graphicx}
\usepackage{mathtools}
\usepackage{ragged2e}
\usepackage{psfrag}

\usepackage{dcolumn}
\usepackage{bm}
\usepackage{graphicx,empheq}
\usepackage{verbatim} 
\usepackage[english]{babel}
\usepackage{hyperref}

\hypersetup{
citecolor=red,
colorlinks=true,
filecolor=red,
linkcolor=blue,
linktocpage=true,
urlcolor=blue
} 
\usepackage[titletoc,toc,title]{appendix}
\numberwithin{equation}{section}
\usepackage{cleveref}
\usepackage{cite}
\usepackage{epsfig}
\usepackage{float}
\usepackage{amsfonts}
\usepackage{enumitem}
\usepackage[font={footnotesize,it}]{caption}

\newcommand\trick[1]{}
\newcommand{\be}{\begin{equation}} 
\newcommand{\ee}{\end{equation}}
\newcommand{\eq}[1]{(\ref{#1})}
\newcommand{\bit}{\begin{itemize}}  \newcommand{\eit}{\end{itemize}}
\newcommand{\ben}{\begin{enumerate}}  \newcommand{\een}{\end{enumerate}}

\newcommand{\rf}[1]{(\ref{#1})}

\def\bd{\begin{document}}
\def\ed{\end{document}}
\def\bea{\begin{eqnarray}}
\def\eea{\end{eqnarray}}
\let\bm=\bibitem

\def\la{\langle}
\def\ra{\rangle}

\def\npb#1#2#3{Nucl. Phys. {\bf{B#1}} #3 (#2)}
\def\plb#1#2#3{Phys. Lett. {\bf{#1B}} #3 (#2)}
\def\prl#1#2#3{Phys. Rev. Lett. {\bf{#1}} #3 (#2)}
\def\prd#1#2#3{Phys. Rev. {D bf{#1}} #3 (#2)}
\def\cmp#1#2#3{Comm. Math. Phys. {\bf{#1}} #3 (#2)}
\def\cqg#1#2#3{Class. Quantum Grav. {\bf{#1}} #3 (#2)}
\def\nppsa#1#2#3{Nucl. Phys. B (Proc. Suppl.) {\bf{#1A}}#3 (#2)}
\def\ap#1#2#3{Ann. of Phys. {\bf{#1}} #3 (#2)}
\def\ijmp#1#2#3{Int. J. Mod. Phys. {\bf{A#1}} #3 (#2)}
\def\rmp#1#2#3{Rev. Mod. Phys. {\bf{#1}} #3 (#2)}
\def\mpla#1#2#3{Mod. Phys. Lett. {\bf A#1} #3 (#2)}
\def\jhep#1#2#3{J. High Energy Phys. {\bf #1} #3 (#2)}
\def\atmp#1#2#3{Adv. Theor. Math. Phys. {\bf #1} #3 (#2)}

\def\sst{\scriptscriptstyle}
\def\thetabar{\bar\theta}
\def\Tr{{\rm Tr}}
\def\one{\mbox{1 \kern-.59em {\rm l}}}

\def\a{\alpha}      \def\da{{\dot\alpha}}  \def\dA{{\dot A}}
\def\b{\beta}       \def\db{{\dot\beta}}
\def\g{\gamma}  \def\G{\Gamma}  \def\dc{{\dot\gamma}}
\def\d{\delta}  \def\D{\Delta}  \def\ddt{\dot\delta}
\def\e{\epsilon}
\def\ve{\varepsilon}
\def\uve{\upvarepsilon}
\def\f{\phi}    \def\F{\Phi}    \def\vvf{\f}
\def\vphi{\varphi}
\def\h{\eta}
\def\k{\kappa}
\def\l{\lambda} \def\L{\Lambda}
\def\m{\mu} \def\n{\nu}
\def\o{\omega}
\def\p{\pi} \def\P{\Pi}
\def\r{\rho}
\def\s{\sigma}  \def\S{\Sigma}
\def\t{\tau}
\def\th{\theta} \def\Th{\Theta} \def\vth{\vartheta}
\def\X{\Xeta}
\def\z{\zeta}

\def\na{\nabla}

\def\cA{{\cal A}} \def\cB{{\cal B}} \def\cC{{\cal C}}
\def\cD{{\cal D}} \def\cE{{\cal E}} \def\cF{{\cal F}}
\def\cG{{\cal G}} \def\cH{{\cal H}} \def\cI{{\cal I}}
\def\cJ{{\mathscr J}} \def\cK{{\cal K}} \def\cL{{\cal L}}
\def\cM{{\cal M}} \def\cN{{\cal N}} \def\cO{{\cal O}}
\def\cP{{\cal P}} \def\cQ{{\cal Q}} \def\cR{{\cal R}}
\def\cS{{\cal S}} \def\cT{{\cal T}} \def\cU{{\cal U}}
\def\cV{{\cal V}} \def\cW{{\cal W}} \def\cX{{\cal X}}
\def\cY{{\cal Y}} \def\cZ{{\cal Z}}
\def\ct{{\cal t}}

\def\ua{\underline{\alpha}}
\def\uc{\underline{\phantom{\alpha}}\!\!\!\gamma}
\def\um{\underline{\mu}}
\def\ud{\underline\delta}
\def\ue{\underline\epsilon}
\def\una{\underline a}\def\unA{\underline A}
\def\unb{\underline b}\def\unB{\underline B}
\def\unc{\underline c}\def\unC{\underline C}
\def\und{\underline d}\def\unD{\underline D}
\def\une{\underline e}\def\unE{\underline E}
\def\unf{\underline{\phantom{e}}\!\!\!\! f}\def\unF{\underline F}
\def\unm{\underline m}\def\unM{{\underline M}}
\def\unn{\underline n}\def\unN{{\underline N}}
\def\unp{\underline{\phantom{a}}\!\!\! p}\def\unP{\underline P}
\def\unq{\underline{\phantom{a}}\!\!\! q}
\def\unQ{\underline{\phantom{A}}\!\!\!\! Q}
\def\unH{\underline{H}}

\def\As {{A \hspace{-6.4pt} \slash}\;}
\def\bs {{b \hspace{-6.4pt} \slash}\;}
\def\Ds {{D \hspace{-6.4pt} \slash}\;}
\def\Gts {{\Gt \hspace{-6.4pt} \slash}\;}
\def\ds {{\del \hspace{-6.4pt} \slash}\;}
\def\ss {{\s \hspace{-6.4pt} \slash}\;}
\def\ks {{ k \hspace{-6.4pt} \slash}\;}
\def\ps {{p \hspace{-6.4pt} \slash}\;}
\def\xs {{x \hspace{-6.4pt} \slash}\;}
\def\pas {{{p_1} \hspace{-6.4pt} \slash}\;}
\def\pbs {{{p_2} \hspace{-6.4pt} \slash}\;}
\def\cFs {{{\cal F} \hspace{-6.4pt} \slash}\;}
\def\Dss {{D \hspace{-7.5pt} \slash}\;}
\def\dss {{\del \hspace{-7.0pt} \slash}\;}

\def\Ah{{\hat{A}}}
\def\Dh{{\hat{D}}}
\def\Gh{{\hat{G}}}
\def\Fh{{\hat{F}}}
\def\Ih{{\hat{I}}}
\def\Jh{{\hat{J}}}
\def\Kh{{\hat{K}}}
\def\Lh{{\hat{L}}}
\def\Ph{{\hat{P}}}
\def\Rh{{\hat{R}}}
\def\Vh{{\hat{V}}}
\def\Xh{{\hat{X}}}

\def\ah{{\hat{\a}}}
\def\bh{{\hat{\b}}}
\def\gh{{\hat{\g}}}
\def\dh{{\hat{\d}}}
\def\rh{{\hat{\r}}}
\def\hh{\hat{h}}
\def\uh{\hat{u}}
\def\xh{\hat{x}}
\def\yh{\hat{y}}
\def\ph{\hat{p}}
\def\xih{\hat{\xi}}
\def\chih{\hat{\chi}}
\def\Psih{\hat{\Psi}}
\def\phih{\hat{\phi}}

\def\psit{\tilde{\psi}}
\def\Psit{\tilde{\Psi}}
\def\Psibt{\tilde{\bar{Psi}}}

\def\lambdat{\tilde {\lambda}}
\def\st{\tilde{\sigma}}

\def\delt{\tilde{\delta}}
\def\Phit{\tilde{\Phi}}
\def\Phitb{\overline{\tilde{Phi}}}
\def\tht{\tilde{\th}}
\def\lt{\tilde{\l}}
\def\chit{\tilde{\chi}}
\def\phit{\tilde{\phi}}

\def\At{\tilde{A}}
\def\Bt{\tilde{B}}
\def\Ct{\tilde{C}}
\def\Dt{\tilde{D}}
\def\Et{\tilde{E}}
\def\Ft{\tilde{F}}
\def\Gt{\tilde{G}}
\def\Ht{\tilde{H}}
\def\It{\tilde{I}}
\def\Jt{\tilde{J}}
\def\Qt{\tilde{Q}}
\def\Rt{\tilde{R}}
\def\Mt{\tilde{M }}
\def\Nt{\tilde{N}}
\def\St{\tilde{S}}
\def\Vt{\tilde{V}}
\def\Xt{\tilde{X}}
\def\at{\tilde{a}}
\def\dt{\tilde{d}}
\def\htt{\tilde{h}}
\def\ft{\tilde{f}}
\def\gt{\tilde{g}}
\def\pt{\tilde{p}}
\def\qt{\tilde{q}}
\def\vt{\tilde{v}}
\def\nt{\tilde{n}}
\def\ut{\tilde{u}}
\def\wt{\tilde{w}}
\def\zt{\tilde{z}}
\def\xt{\tilde{x}}
\def\yt{\tilde{y}}
\def\Psit{\tilde{\Psi}}
\def\vphit{\tilde{\varphi}}
\def\tD{\tilde{\D}}

\def\eb{\bar{\epsilon}}
\def\delb{\bar{\partial}}
\def\thb{\bar{\theta}}
\def\mub{\bar{\mu}}
\def\lamb{\bar{\l}}
\def\psib{\bar{\psi}}
\def\sb{\bar{\sigma}}
\def\xib{\bar{\xi}}
\def\chib{\bar{\chi}}

\def\Psib{\bar{\Psi}}
\def\Phib{\bar{\Phi}}
\def\Lamb{\bar{\Lambda}}
\def\Sb{{\overline \Sigma}}
\def\cb{\bar{c}}
\def\hb{\bar{h}}
\def\qb{\bar{q}}
\def\wb{\bar{w}}
\def\ub{\bar{u}}
\def\zb{{\bar{z}}}
\def\Hb{\bar{H}}
\def\Qb{{\bar Q}}
\def\Omegab{\overline{\Omega}}
\def\ob{\overline{\omega}}

\def\Ab{{\overline A}} \def\Bb{{\overline B}} \def\Cb{{\overline C}}
\def\Db{{\overline D}} \def\Eb{{\overline E}} \def\Fb{{\overline F}}
\def\Gb{{\overline G}}
\def\Ib{{\overline I}}
\def\Jb{{\overline J}} \def\Kb{{\overline K}} \def\Lb{{\overline L}}
\def\Mb{{\overline M}} \def\Nb{{\overline N}} \def\Ob{{\overline O}}
\def\Pb{{\overline P}}  \def\Rb{{\overline R}}
 \def\Tb{{\overline T}} \def\Ub{{\overline U}}
\def\Vb{{\overline V}} \def\Wb{{\overline W}} \def\Xb{{\overline X}}
\def\Yb{{\overline Y}} \def\Zb{{\overline Z}}

\def\fb{{\overline f}}
\def\gb{{\overline g}}
\def\nb{{\overline n}}
\def\mb{{\overline m}}
\def\lb{{\overline l}}
\def\yb{{\overline y}}

\def\ldel{{\overleftarrow{\del}}}
\def\rdel{{\overrightarrow{\del}}}
\def\ldeldel{{\overleftarrow{\del^2}}}
\def\rdeldel{{\overrightarrow{\del^2}}}
\def\ldelb{{\overleftarrow{\bar{\del}}}}
\def\rdelb{{\overrightarrow{\bar{\del}}}}

\def\ba{{\bf a}}
\def\bk{{\bf k}}
\def\bl{{\bf l}}
\def\bp{{\bf p}}
\def\bq{{\bf q}}
\def\br{{\bf r}}
\def\bt{{\bf t}}
\def\bu{{\bf u}}
\def\bv{{\bf v}}
\def\bx{{\bf x}}
\def\by{{\bf y}}
\def\bA{{\bf A}}
\def\bR{{\bf R}}
\def\bV{{\bf V}}

\def\bz{{\boldsymbol{\zeta}}}

\def\bone{{\bf 1}}

\def\va{{\vec a}}
\def\vk{{\vec k}}
\def\vp{{\vec p}}
\def\vq{{\vec q}}
\def\vx{{\vec x}}
\def\vy{{\vec y}}
\def\vu{{\vec u}}
\def\vv{{\vec v}}
\def \vH{{\vec H}}
\def \vg{{\vec g}}

\def\vs{{\vec \sigma}}
\def\vtau{{\vec \tau}}

\newcommand{\ov}[1]{\overrightarrow{#1}}

\def\frA{\mathfrak{A}}
\def\frB{\mathfrak{B}}
\def\frC{\mathfrak{C}}
\def\frD{\mathfrak{D}}
\def\frE{\mathfrak{E}}
\def\frF{\mathfrak{F}}
\def\frG{\mathfrak{G}}
\def\frH{\mathfrak{H}}
\def\frM{\mathfrak{M}}
\def\frN{\mathfrak{N}}
\def\frR{\mathfrak{R}}
\def\frW{\mathfrak{W}}

\def\fra{\mathfrak{a}}
\def\frb{\mathfrak{b}}
\def\frf{\mathfrak{f}}
\def\frg{\mathfrak{g}}
\def\frh{\mathfrak{h}}
\def\frl{\mathfrak{l}}
\def\frs{\mathfrak{s}}
\def\fri{\mathfrak{i}}
\def\frj{\mathfrak{j}}

\def\ma{\mathfrak{a}}
\def\mg{\mathfrak{g}}
\def\mh{\mathfrak{h}}
\def\mR{\mathfrak{R}}
\def\mN{\mathfrak{N}}

\newcommand{\nn}{{\nonumber}}

\def\d{\delta}\def\D{\Delta}\def\ddt{\dot\delta}

\def\pa{\partial} \def\del{\partial}
\def\xx{\times}
\def\uno{\mbox{1 \kern-.59em {\rm l}}}

\def\trp{^{\top}}
\def\inv{^{-1}}
\def\dag{\dagger}
\def\pr{^{\prime}}

\def\rar{\rightarrow}
\def\lar{\leftarrow}
\def\lrar{\leftrightarrow}

\newcommand{\0}{\,\!}      
\def\one{1\!\!1\,\,}
\def\im{\imath}
\def\jm{\jmath}

\newcommand{\tr}{\mbox{tr}}
\newcommand{\slsh}[1]{/ \!\!\!\! #1}

\newcommand{\1}{\mbox{1}\hspace{-0.25em}\mbox{l}}

\def\vac{|0\rangle}
\def\lvac{\langle 0|}

\def\hlf{\frac{1}{2}}
\def\ove#1{\frac{1}{#1}}
\newcommand{\hot}[1]{\frac{#1}{2}}

\def\Box{\square}
\def\CC {\mathbb{C}}
\def\FF {\mathbb{F}}
\def\RR{\mathbb{R}}
\def\NN{\mathbb{N}}
\def\ZZ{\mathbb{Z}}
\def\bb#1{{\bf #1}}
\def\bcomment#1{}
\def\bfhat#1{{\bf \hat{#1}}}
\def\VEV#1{\left\langle #1\right\rangle}

\newcommand{\ex}[1]{{\rm e}^{#1}} \def\ii{{\rm i}}

\newcommand{\lrbrk}[1]{\left(#1\right)}
\newcommand{\lrsbrk}[1]{\left[#1\right]}
\newcommand{\sfrac}[2]{{\textstyle\frac{#1}{#2}}}

\def\stw{{\sqrt{2}}}

\def\rf {{\rm f}}
\def\ri {{\rm i}}
\def\rj {{\rm j}}
\def\rn {{\rm n}}
\def\rk {{\rm k}}
\def\rl {{\rm l}}
\def\rr {{\rm r}}
\def\rs {{\scriptscriptstyle \rm S}}
\def\rt {{\scriptscriptstyle \rm T}}

\def\rQ {{\scriptscriptstyle \rm \cQ}}
\def\rR {{\scriptscriptstyle \rm \cR}}

\def\cQb{{\cal \Qb}}
\def\cRb{{\cal \Rb}}
\def\cWb{{\cal \Wb}}

\def\fd {{\rm N}}
\def\afd {{\overline{\rm N}}}

\def \II {I\hspace{-.1em}I\hspace{.1em}}
\def \IIA {\mbox{\II A\hspace{.2em}}}
\def \IIB {\mbox{\II B\hspace{.2em}}}
\def \gs {g^s}
\def \ls {\lambda^s}

\def \I {{\cal I}}
\def \qs {q\hspace{-.53em}/\hspace{.15em}}
\def \ks {k\hspace{-.53em}/\hspace{.15em}}
\def \YM {{\mbox{\tiny YM}}}
\def \gym {g_{\YM}}

\def \Lc {\L_c}
\def\IR{\relax{\rm I\kern-.18em R}}
\def \id {{\bf 1}}

\def\cci{\ell}
\def\ccj{\ell'}

\def\bbq{\pmb{q}}
\def\bom{\pmb{\o}}
\def\bJ{\pmb{J}}
\def\bM{\pmb{M}}
\def\bB{\pmb{B}}
\def\bn{\pmb{n}}
\def\bE{\pmb{E}}

\newcommand{\rrr}[1]{\vskip 0.2cm \noindent{\bf #1} ---}

\long\def\symbolfootnote[#1]#2{\begingroup%
\def\thefootnote{\fnsymbol{footnote}}\footnote[#1]{#2}\endgroup}
\long\def\RemarkBox#1{\begin{flushleft}\fbox{\begin{minipage}
{17.5cm}{\bf Remark:} ~#1\end{minipage}}\end{flushleft}}
\newcommand{\aei}{\it Max Planck Institute for Gravitational Physics
(Albert Einstein Institute)\\ Am M\"uhlenberg 1, 14476 Golm,
Germany}

\newcommand{\nthu}{{\it Department of Physics, National Tsing-Hua
  University,
  Hsinchu 30013, Taiwan}}

\newcommand{\ctc}{{\it
Center of Theory and Computation, 
National Tsing-Hua University, Hsinchu 30013, Taiwan}}

\newcommand{\ictsustc}{{\it Interdisciplinary Center for Theoretical Study,
University of Science and Technology of China,\\
Hefei, Anhui 230026, People's Republic of China}}

\newcommand{\sysu}{{\it School of Physics and Astronomy, Sun Yat-Sen
    University, Zhuhai 519082, China}}

\newcommand{\ncts}{{\it
    National Center for Theoretical Sciences, Taipei 10617, Taiwan}}

\begin{document}

\begin{center}
~\vspace{20pt}
  
\thispagestyle{empty}
              {\Large \bf 
Tunneling of Bell Particles, Page Curve and\\ Black Hole Information}
\vspace{25pt}
 
Chong-Sun Chu ${}^{2,3,4}$\symbolfootnote[1]{Email:~\sf
  cschu@phys.nthu.edu.tw}, Rong-Xin Miao
${}^1$\symbolfootnote[2]{Email:~\sf  miaorx@mail.sysu.edu.cn}

\vspace{10pt}${{}^{1}}$\sysu \symbolfootnote[3]{All the Institutes of authors
  contribute equally to this work, the order of Institutes is adjusted
  for the assessment policy of SYSU.}\\
\vspace{5pt}${{}^{2}}$\ctc\\
\vspace{5pt}${{}^{3}}$\nthu\\
\vspace{5pt}${{}^{4}}$\ncts

\vspace{1cm}


\begin{abstract}
  We propose that the quantum states of black hole responsible for the
  Bekenstein-Hawking entropy are given by a thin shell of Bell
  particles located at the region just underneath the horizon. We
  argue that the configuration can be stabilized by a new kind of
  degeneracy pressure which is suggested by
  a noncommutative geometry
  in the interior of the black hole. Black hole singularity is avoided.
  We utilize the work of Parikh and Wilczek \cite{Parikh:1999mf} to
  include the effect of tunneling on the Bell particles.  We show that
  partially tunneled Bell particles give the Page curve
  of
  Hawking radiation, and the entirety of information initially stored
  in the black hole is returned to the outside via the Hawking
  radiation.  In view of entropic force, the location of these Bell
  states is naturally related to the island and the quantum extremal
  surface.

\end{abstract}

\end{center}

\newpage
\setcounter{footnote}{0}
\setcounter{page}{1}

\tableofcontents

\setcounter{footnote}{0}

\section{Introduction}
There are mounting theoretical
evidences that 
a black hole obeys the first law of
thermodynamics with an entropy
\be \label{BH}
S_{\rm BH} = \frac{A}{4G}
\ee
and
a temperature
$T =1/8 \pi M$
due to a thermal Hawking radiation \cite{Hawking:1975vcx}.
Despite remarkable progress that has been
made in microstate counting 
\cite{Strominger:1996sh}, 
it is still not known what is
the nature of the gravitational degrees of freedom that are being 
counted by \eq{BH}.

Another outstanding problem of the black hole
is the information problem
\cite{Hawking:1976ra}.
Consider a black hole form from a pure state. Assuming
that the Hawking radiation is thermal, then the entanglement entropy
outside the black hole increases monotonically for the entire course
of life of the black hole. This  {\it Hawking curve} 
violates the fundamental unitarity principle of quantum mechanics:
the fine-grained entanglement entropy should not exceed the
coarse-grained black hole entropy.
On the other hand, Page argued that \cite{Page:1993wv,Page:2013dx}
if the black hole evolution process is unitary, then
the total system of black hole and radiation
must go back to a pure state at the end of the evaporation.
As a result, unitarity of quantum black hole
requires the following properties for the
Hawking radiation:
1. Page curve:
  the entanglement entropy of Hawking radiation should initially
  rise until the so called Page time $t_P$ when it starts to drop down to zero
  as the black hole  completely vanishes. 
2. Recovery of information: at the end of the black hole life,
  one must be able to recover from the Hawking
  radiation all the information of the  initial pure state. 
  It was Page's remarkable insight that the Page curve behavior
  of the Hawking  radiation,
  due to its accessibility to external observers,
  can be especially useful as a
  decisive
  criteria for  unitary theory of quantum gravity.
  
  Recently, by using a fully quantum form \cite{Engelhardt:2014gca}
of the Ryu-Takayanagi entropy formula  \cite{Ryu:2006bv,Hubeny:2007xt},
the entanglement entropy
of the Hawking radiation was computed and shown to obey the Page curve
\cite{Penington:2019npb,Almheiri:2019psf,Almheiri:2019hni,Almheiri:2019hni}.
  Central to this analysis is the emergence of island, 
  regions of spacetime  
  outside or inside the horizon
  that are completely disconnected and
spacelike separated from the region of the Hawking radiation. 
From  the semi-classical point of view,
island originated from the  wormhole saddle in the replica path
integral for the entanglement entropy \cite{Almheiri:2019qdq}.
We note that the island 
 is located
just beneath the horizon for evaporating Schwarzschild black hole
\cite{Penington:2019npb,Almheiri:2019psf,Almheiri:2019hni}.
It would be interesting to understand the connection between the island picture and the tunneling 
picture for the evolution of black hole that we develop here.

That the Page curve is obtained
gives confidence that the AdS/CFT correspondence and
the generalized entropy formula 
constitutes a credible quantum gravity
framework to study 
black holes. Yet, it still leaves many questions unanswered.
For example, it is believed that the Bekenstein-Hawking entropy \eq{BH} is
coarse-grained in nature
\cite{Engelhardt:2017aux,Almheiri:2020cfm}. From this perspective, 
what role does island play in coarse graining the
initial pure state? How does it help in
returning the information to the environment?
The main motivations of this work have been to understand
better  
the origin of the Page curve and the return of information
in terms of explicit spacetime quantum mechanical processes.

\section{Quantum pressure and interior distribution of degrees of freedom}
General relativity predicts that any object
that has collapsed beyond a certain
point would form a black hole, inside which there is a singularity. For a
spherical symmetric black hole, the spacetime
is given by the Schwarzschild
metric, a vacuum solution ($T_{\m\n} =0$) of the Einstein equation, and the
singularity at the origin represents a place where matter is compressed
infinitely and the classical description of spacetime breaks down. 
In the absence of a new source of pressure to counterbalance the
gravitational pull, the collapse of black hole to a singularity
is unavoidable. However it should be noted that the
singularity theorem \cite{Penrose:1964wq} is based on the canonical
assumptions of general relativity
and certain energy conditions on the matter. 
It has been widely speculated that quantized gravity would resolve the
singularity and provide a complete description of the
interior of black hole. The singularity may also be resolved in the form of
wormhole due to the existence of quantum energy.
Given little is known inside the black hole,  we consider in this
paper the possibility of a novel kind of 
degeneracy pressure for the degrees of freedom in the interior of
a black hole that can counter-balance 
the gravitational collapse.

\subsection{Degeneracy Pressure of Neutron Star}
To
get motivated, let us briefly review the collapse process of
a compact star. Let $M$ be the mass and $\r$ be the radius of the star.
The gravitational energy
is of the order of $E_g \sim - GM^2/\r$. This give rises to a
gravitational pressure $P_g = - \del E_g /\del V$
\be \label{Pg}
P_g \sim - G M^2/\r^4.
\ee
The collapse would be halted if the gravitational pressure
is balanced out by some
pressure produced by the star matter.
The strongest known matter pressure is the neutron degeneracy pressure.
It is useful to review it's origin. Consider a gas of free
particles with a 
dispersion relation  of the form $E = d_0 p^\b$, where $E$ is the kinetic
energy, and  $d_0$, $\b$ are constants. This covers the non-relativistic case
$E = p^2/2m$, fully relativistic case $E =p$, as well as possible quantum
deformed dispersion relations (QDR) with $\b \neq 1,2$. 
The density of states is
$g(E) dE = {g_s V 4 \pi p^2 dp}/{(2 \pi \hbar)^3} $, $g_s = 2s+1$ is a
spin factor,
then takes the form
\be \label{QDR}
g(E) = c_0 V E^\a,
\ee
where $\a= {3}/{\b} -1 $ and $c_0$ is a constant. For a fully degenerate
system of fermions,
the occupation number $n(E)$  is given by
\be
n(E) = \begin{cases}
  1, & E \leq \mu, \\
  0, & E > \mu,
  \end{cases}
  \ee
where $\mu$ is the Fermi  level at zero temperature. 
This gives the total number of
particles $N = \int d E g(E) n(E)$ and  the total kinetic energy $U = \int dE
g(E) n(E) E $ as
\be \label{N}
N= \frac{c_0 V}{\a+1} \m^{\a+1},
\ee
\be
\label{U}
U = \frac{\a+1}{\a+2} N \mu. 
\ee
As a result, the matter pressure $P_m = - (\del U /\del V)_N$ reads
\be\label{Pm-mu}
P_m \approx  - N \frac{\del \mu}{\del V},
\ee
up to an order 1 
proportional constant $ \frac{\a+1}{\a+2}$. 
Therefore, the degeneracy pressure is determined by the volume dependence of
the Fermi level $\mu$. For non-relativistic neutron star, $\a=1/2$. \eq{N}
gives
$\mu = \frac{1}{2m}(6 \pi^2 N/V)^{2/3}$
and
\be \label{Pm-neutron}
P_m \sim \frac{ N^{5/3}\hbar^2}{m \r^5},
\ee
where $m$ is the mass of neutron. 
The collapse is halted if $P_g + P_m =0$ and this is possible if
the mass limit  $M \lesssim 2 M_{\odot} $ is satisfied.
Physically, the degeneracy pressure \eq{Pm-neutron} which stabilize 
the neutron star arises from the fact that all the available energy
levels of the system are filled such that no further addition of
states is possible.

\subsection{Black Hole Interior}
Back to the case of black hole. The fact that black hole has an entropy and a
temperature suggests that black hole is not a classical vacuum as
described in general relativity, but a nontrivial quantum system of
microstates. Obviously these cannot be elementary particles of the
standard model since ordinary energy-momentum cannot provide a pressure
strong enough to withstand the collapse of gravity. Instead, these quanta
should have an universal nature that is independent of the matter
that has been collapsed to form the black hole. It then seems natural
that these are the elementary quanta of black hole spacetime itself.
In the following,  we will ignore their mutual
interaction and model the quantum black hole as a gas of free particles.
As these microstates arises from the quantization within the compact interior
of black hole, it is natural that they carry an average energy 
of the order  $\mu \sim 1/\rho$. We will now see that this simple assumption
tells us something interesting about the nature of these microstates.

The area dependence of the Bekenstein-Hawking entropy has led to the
formulation of the holographic principle
\cite{tHooft:1993dmi,Susskind:1994vu} and has been the chief guiding
route to the understanding of quantum gravity.
For a microcanonical ensemble, the entropy takes the form $S = N \ln
\Omega$ where $N$ is the number of degrees of freedom and $\Omega$ is
the phase space volume available to each individual degrees of
freedom.
Normally the entropy of a many bodies continuum quantum system is
divergent since there is an infinite volume of phase space available
to each degree of freedom.  That the black hole entropy \eq{BH} is
finite means that not just the degrees of freedom making up the black
hole is finite, but also the phase space volume available to each
degree of freedom is finite.
In fact, in the holographic picture of \cite{Susskind:1994vu},
there is a two states spin system (hence $\Omega =2$) associated with
each site of a two dimensional lattice of quanta (called ``partons'')
living on the horizon area.
Now in our model, using the fact that the energy of the system
must be equal to the mass $M$ of the black hole,
we obtain immediately from \eq{U}
\be
N \sim M \r \sim A/G
\ee
and
the entropy \eq{BH} is reproduced  if the
number of available states to each particle is a finite constant.
This simple interpretation of the formula \eq{BH} suggests that the 
the entirety
of the black hole microstates are distributed over a thin shell
near the horizon. In this case, the total energy of the system is 
$E_m \sim N \hbar c /\r$ and \eq{Pm-mu} gives the degeneracy pressure,
\be \label{Pm}
P_m \sim N\hbar/\r^4,
\ee
which cancels  that \eq{Pg} of  gravity
up to  numerical factor of order 1.
This simple analysis suggests that
the collapsed matter can indeed be stabilized
just underneath the horizon by the degeneracy pressure.

So what is the reason for this exclusion of states? One possible origin
of this is noncommutative geometry:
that the interior of black hole is in fact described by a quantized
space with an uncertainty relation
\be \label{ur}
\D V \gtrsim l_P^3.
\ee
In this case, the number of states that is available in the
region of space underneath the horizon and with a thickness
$w\sim l_P$ is given by $N \sim \r^2/l_P^2$.
It explains \eq{BH}.
We note that in this picture, the horizon is at the junction
between the exterior commutative
geometry and the interior
quantum geometry, and the degeneracy pressure
may be thought of as some kind of interface pressure.

It is amazing that the quantum extremal surface
for Schwarzschild black hole is also located just underneath the horizon.
One may speculate that this is how the
quantum space and the collapsed matter are
described as island and quantum extreme surface in
holography.
Recall the location of island is determined by minimizing the
generalized entropy with contributions from both gravity and matters,
i.e. $\delta S_{\text{gen}}=0$ \cite{Almheiri:2020cfm}. In the
viewpoint of entropic force \cite{Verlinde:2010hp}, this means the
quantum extreme surface is the force balance surface
$F=TdS_{\text{gen}}/d\rho=0$. It is consistent with our discussions of
quantum pressure
$P_g+P_m=F/A=0$, where $P_g$ and $P_m$ originated
from the entropy of gravity and matters, respectively.

Summarizing, we propose that,
instead of a singularity,
the interior of black hole is described by a
quantum 
geometry with the uncertainty relation \eq{ur}.
The resulting maximal
occupancy of states provides a novel degeneracy pressure which
stabilizes the collapsed matter at a thin region (thickness $w \sim l_P$)
right beneath the horizon.

\section{Black hole interior as
  thin shell of Bell particles}
Consider a black hole form in flat space
from a pure state and decays under Hawking radiation.
If unitarity of
quantum mechanics is not violated in the black hole formation process,
then
the black hole interior
and  exterior together is in a pure state. 
Due to our ignorance of the interior, this give rises to a number
$S_{\rm BH}$ of coarse-grained states with entropy \eq{BH}.
By definition, the knowledge of these states together with the knowledge
of the exterior constitutes a pure state. Eventually,
the black hole is exhausted by the Hawking radiation. If we assume that
unitarity is preserved throughout, then the final state of the Hawking
radiation cannot be purely thermal, but it must encompass the information
contained in the initial set of coarse-grained states so that,
together with the outside information, a pure state
can be reconstructed.

It has been suggested that
some form of nonlocality is needed to resolve the black hole
information problem (see \cite{Giddings:2006sj} for a review).  
As quantum nonlocality is best captured by entanglement, it seems natural
to consider the coarse-grained degrees of freedom to be rich in entanglement
content. 
We propose in this paper that the black hole interior degrees of freedom
are given by maximally entangled Bell pairs of
particles
localized on the
internal side of the horizon.
Is this  observable to the outside? Our main idea is that 
tunneling can reveal the interior
information of black hole to the outside world.

Let us briefly comment on the Hawking radiation. In addition to
being an effect of QFT in curved spacetime,
a particularly transparent understanding of
the origin of Hawking radiation is the
quantum mechanical picture of Parikh and Wilczek \cite{Parikh:1999mf},
where the interior of black hole
is considered to be in a vacuum state and the effect of  tunneling on
the virtually created particles resulted in the Hawking radiation.
For an outgoing particle with energy $\o$ created just inside the horizon,
they found that it can travels cross the horizon and results in
a nonvanishing imaginary part for the
particle action:
$
{\rm Im} S = 4 \pi \o (M-\frac{\o}{2}).
$
This corresponds to a tunneling process with the tunneling rate 
$ \G = \G_0 e^{-2 {\rm Im} S} = \G_0 e^{-8\pi \o (M-\frac{\o}{2})}$,
where $\G_0$ 
is a prefactor which can be computed from a more detailed knowledge
of the dynamics.
The leading
exponential $\o$ dependence in $\G$ registers a
Boltzmann 
thermal distribution with the Hawking
temperature $T$.
The $\o^2$ term
is a back reaction term, suggesting that the spectrum is
slightly deviated from the thermal one.
In addition,
pair creation  outside the horizon was also
considered.
In this so called anti-particle channel, the anti-particle follows a time
reversed ingoing geodesic crosses the horizon and also makes
contribution to the Hawking radiation. 
Note
that while the original
analysis of \cite{Parikh:1999mf} is for massless particles,
it
applies to massive particles as well and the tunneling rate is the same.

Although this picture of tunneling explains very physically
the existence of Hawking radiation and its temperature,
it also
leads to the Hawking curve.
To see this, 
consider a virtual pair created
in the interior side of the horizon (particle channel).
Since the pair was created entangled,
an entanglement between the black hole and the Hawking radiation is
created when the
particle tunnels through and the
anti-particle got absorbed by the black hole.
The rate of increase of the entanglement entropy of the Hawking radiation
is given by $\a (\o) A$, where
$\o$ is the energy of the particle
and $\a(\o)$ is
the creation rate per unit area of
virtual pairs on the interior side of horizon
times
the tunneling probability.
Similarly,
there is a contribution 
$\b (\o) A$
from the anti-particle channel,
where $\b(\o)$ is the creation rate of
virtual pairs on the exterior side of horizon
times the tunneling probability.
As $\a, \b$ are positive, 
the entanglement entropy of
Hawking radiation increases monotonically. This Hawking curve violates
the unitarity of quantum mechanics.

We note that the main difference between the consideration of
\cite{Parikh:1999mf} and ours is in the assumption
concerning the black hole interior.
Instead of a vacuum as considered in \cite{Parikh:1999mf},
we propose that there is a thin shell of Bell pairs
sitting right beneath the horizon
and we expect the
mechanism for their tunneling will be different from that
of \cite{Parikh:1999mf}. 
Although we do not currently have the
technology of quantum
gravity to determine the tunneling rate, 
nevertheless a generic analysis can be performed and 
our following conclusion 
will not depend
on the detailed form  of the tunneling rate.
We will next show how  
tunneling in our model affects the  Bell pairs
and leads to the Page curve
and the recovery of the full entanglement
content of the coarse-grained states in  late time of the Hawking radiation.

We emphasis that our thin shell is not a thin shell of material
particles, but a thin shell of quantum bits of space.  Incidentally,
an interesting model of black hole evaporation has been proposed
\cite{Kawai:2013mda} where a thin shell of collapsed matter is
featured.  We suspect that matter and spacetime will be unified in
quantum gravity. It is an intriguing question to understand how a thin
shell of collapsed matter would turn into a thin shell of quantum bits
of space.  See also \cite{Kawai:2015uya,VanRaamsdonk:2013sza,
  Akhmedov:2015xwa, Ho:2015vga, Good:2016atu,
  Baccetti:2016lsb,Mann:2021mnc,Ho:2022gpg} for more interesting
discussion concerning Hawking radiation in semiclassical model of
black hole.

\section{Tunneling and entanglement swapping of Bell pairs}
Due to their close proximity to the horizon,
both particles of the Bell pairs can tunnel and
leave the black hole as Hawking radiation.
In addition, the Bell particles also have an interesting
effect of {\it entanglement swapping}. 
Consider the particle channel, the anti-particle $\bar{p}$
that is left behind can get annihilated by a particle $b_2$
of one of the Bell pairs.
As a result, the particle $p$
becomes entangled with the other
partner $b_1$ of the Bell pair, and entanglement is swapped from
the Bell pair and the virtual pair to one between the
Hawking particle and the
remaining Bell particle inside the black hole.
Similarly, 
entanglement swapping occurs in the anti-particle channel.
See  Figure 1.
\begin{figure}
  \label{swap}
  \centering
      \includegraphics[width=0.67\linewidth]{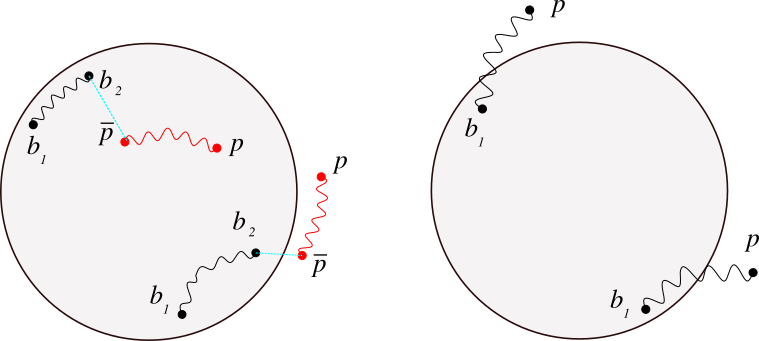}
      \caption{Entanglement swapping induced by virtual pairs}
\end{figure}

Both of these processes will not only alter the entropy content of the black
hole, but also introduce an 
entanglement entropy for the Hawking radiation due to the partially escaped 
entangled pairs.
To study the dynamics of the
entangled pairs of particles,
let us
denote the energies of the particles of each
entangled pair by
$0\leq \o_1, \o_2 \leq M_0$, where $M_0$ is the original mass of the black hole.
Let $n_1(\o_1,\o_2,t) d\o_1 d\o_2$
be the number
of
entangled pairs that are located entirely inside the horizon and
have energies within the intervals $(\o_1,\o_1+d \o_1)$ and
$(\o_2,\o_2+d \o_2)$. 
Similarly, let us denote by $n_2(\o_1,\o_2,t) d\o_1 d\o_2$
the number of
entangled pairs that has the $\o_1$-particle 
inside the horizon and $\o_2$-particle outside, and
 $n_3(\o_1,\o_2,t) d\o_1 d\o_2$
the number of
entangled pairs  that are located entirely outside the horizon.
In order to avoid over counting,
the domain of energies for $n_1(\o_1,\o_2), n_3(\o_1,\o_3)$  is given by
$D := \{ (\o_1,\o_2) |\;  0\leq \o_1 \leq \o_2 \leq M_0 \}$.
As for $n_2(\o_1,\o_2,t)$, the
first (resp. second) argument $\o_1$ (resp. $\o_2$)
refers to the energy of the particle that
is inside (resp. outside)
the horizon. There is no constraint on the size of
$\o_1, \o_2$ and the
domain for $n_2$ is given by
$ D' := \{ (\o_1,\o_2) | \;  0\leq \o_1, \o_2 \leq M_0 \}$.
See Figure 2. 
\begin{figure}
  \label{bell}
  \centering
\includegraphics[width=0.32\linewidth]{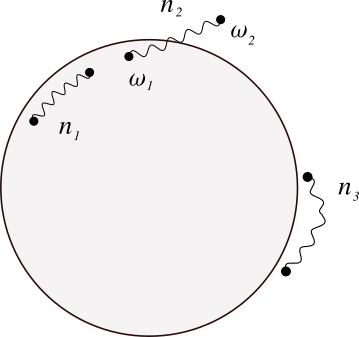}
\caption{Tunneling of
entangled particles}
\end{figure}

Taking into account of tunneling
of the Bell particles and
entanglement swapping, it is easy to obtain
\begin{align}
  \frac{\del n_1}{\del t} &= - (\G_1  + \G_2)\, n_1-(\d_1+\d_2)A,
  \; \hspace{1.3cm}
  \label{n1}
\end{align}
\vspace*{-4ex}
  \begin{subequations} \label{n2}
\begin{align}
    \frac{\del n_2}{\del t} &=  \G_2 n_1 -\G_1 n_2 +\d_1 A, \quad
\mbox{for $\o_1\leq \o_2$}, \label{n2a}\\
\frac{\del n_2}{\del t} &=  \G_2 \nt_1 -\G_1 n_2  + \d_1 A,
\quad \mbox{for $\o_1 > \o_2$},
\label{n2b}
\end{align}
  \end{subequations}
\vspace*{-4ex}
\begin{align}
  \frac{\del n_3}{\del t} &= \G_1 n_2 + \G_2 \nt_2,
  \; \hspace{3.6cm} \label{n3}
\end{align}
where $\G_a := \Gamma(\o_a,M(t))$ ($a=1,2$)
is the tunneling rate for a particle of
energy $\o_a$ from a
black
hole of mass $M(t)$.
$\a_a :=\a(\o_a,M(t))$ (resp. $\b_a :=\b(\o_a,M(t))$)
is the production rate per unit area of the conventional
vacuum created Hawking radiation in the anti-particle channel (resp. particle
channel), and $\d_a:=\a_a+\b_a$ is the total
production rate from both channels.
Here $n_i = n_i(\o_1,\o_2,t)$,
while $\nt_{1,2} := n_{1,2} (\o_2,\o_1,t)$ has its arguments reversed.
Physically, the
$n_i$ terms and the $A$ terms
on the RHS of \eq{n1} (resp. \eq{n2} or \eq{n3}) represent
the effect of direct tunneling and quantum swapping 
on the
entangled pairs that are entirely inside the black hole (resp.
partially inside or entirely outside the black hole).

The total number of each types of
entangled pairs is
\be
N_i (t) = \int_{D_i} d\o_1 d\o_2 n_i (\o_1,\o_2,t),\quad i=1,2,3,
\ee
where, as explained above, $D_1 =D_3 = D$ and $D_2 =D'$.
Physically,
$(2N_1 +N_2)$ 
represent
the amount of coarse-grained entropy of the black hole at time $t$,
the $N_2$-part of which is entangled with the outside observer
and gives the entanglement entropy of Hawking radiation. As for
$N_3$, it represents the amount of entanglement information contained in the
Hawking radiation.
In our model, the mass of the black hole is given by
\be \label{Mt}
M(t) = \int d\o_1 d\o_2 \left[
  (\o_1 + \o_2) n_1 + \o_1 n_2
\right]
  \ee
  where the appropriate domains of integration is understood. In a
  consistent analysis, \eq{Mt} will have to be considered together with
  \eq{n1} - \eq{n3}. This is quite a complicated system to solve.
  It turns out that without assuming the form of $\G(\o,M)$ and
  $\d(\o,M)$, and without solving the system,
  one can immediately show that
  the Hawking radiation
  obeys the Page curve and that the complete information of the black hole
  is returned.

Let us consider a black hole formed at $t=0$, implying
the conditions $n_2 =n_3 =0$ initially.
Consider first \eq{n1}. As
the RHS of \eq{n1} is 
non-positive, 
$n_1$ will continue to decrease until
it reaches zero, where also $\del n_1/\del t =0$.
Denote this time as $t_{1,\o_a}$ and define
$t_1 : = {\rm Supp}_{(\o_1,\o_2)\in D} (t_{1,\o_a})$.
We have
\be
N_1=0 \;\;\; \mbox{for $t\geq t_1$}.
\ee
Next  consider \eq{n2}. Starting with the initial conditions
$n_2=0$ and $n_1 >0$, $n_2$ is set to increase initially.
As time progress, there will be a cross over time  where the RHS of
\eq{n2} vanishes and turns negative subsequently. Then
$n_2$ will continue to decrease until it reaches zero
at some time $t_{2,\o_a}$.
Define
$ t_E : = {\rm Supp}_{(\o_1,\o_2)\in D'} (t_{2,\o_a})$,
we have
\be
N_2 =0 \;\;\; \mbox{for $t\geq t_E$}.
\ee
Finally  consider \eq{n3}. It is clear that $n_3$ increases monotonically
until $n_2$ reaches zero at $t=t_{2,\o_a}$, then it remains a constant.
Therefore we have
\be
\mbox{$N_3 = N_{3f}$ a constant}\;\;\; \mbox{for $t \geq t_E$}.
\ee
It is clear that $t_1$ is when the black hole's
degrees
of freedom become completely entangled with the environment
and
$t_E$ is the end time that the black hole dies.

Note that as
$N_2$ starts zero at $t=0$
and ends at zero again at $t=t_E$, it must
follow an inverted V-shape curve
and reaches a maximum at some
intermediate time $t_P$. This is generic and we obtain 
the Page
curve for the Hawking radiation, with $t_P$
being the Page time. 
By expressing
$N_2$ as integrals over the domain $D$, it is easy to establish
the conservation equation:
\be \label{cons}
\frac{d}{dt} (N_1+ N_2 + N_3) =0.
\ee
It is remarkable that
this leads immediately to
\be
N_{3f} = N_{10},
\ee
meaning that all the entanglement
information originally stored in the
Bell pairs are
returned to the exterior observer via the Hawking radiation.
We note that it is crucial in
our model to include the thin shell of Bell particles
to start with and allow the tunneling of them.
Without
these,
our equations \eq{n1}-\eq{n3} will reduce to the equation
$\del n_2/\del t = \d_1 A$
as in the conventional tunneling model of Hawking radiation.
There
would then be the Hawking curve and it
would not possible to return information
via $n_3$. 
In the
appendix,
we justify that $t_E$ is finite in our model.
We also show the explicit time dependence of the
$n_i$'s for some typical values of the energies, which confirm the generic
behavior
discussed here.

Note that in our model, the Hawking radiation
is non-thermal and contains correlations that
are
necessarily highly nonlocal as
they could
arise
from tunneling process that occurs at very different times
over the life of the black hole.
We remark that an infalling observer 
will encounter the thin shell of entangled particles
located underneath the horizon
and get assimilated there as new Bell particles. 
In this sense the
shell of
Bell particles acts
like a firewall \cite{Almheiri:2012rt}. 
We have considered and
proposed that the
wave function of the
thin shell matter can
be written in terms
of 2-qubit Bell states. It is interesting to understand
if this is true and
how the firewall makes it.

{\it Note added:}  Since it's original submission, 
progress has been made in future developing the idea 
suggested in this paper.
In \cite{Chu:2023mqi}, a bottom-up approach to quantum black hole and
quantum gravity was advocated. It was found that in order for a quantum
mechanical system of Fermi degrees of freedom to emulate a black hole
(i.e. reproducing the size and shape of a black hole, and the
Bekenstein-Hawking entropy by an microstate counting), it is
sufficient for the underlying Hamiltonian to admit a Fermi sea with an
uniformly density of states. More recently, a proposal of quantum
gravity as a certain quantum mechanics of noncommutative spaces was
put forwarded \cite{Chu:2024qil,Chu:2024edh}. It was shown that the
theory admits noncommutative fuzzy geometries as solutions whose
energetic properties (energy, size, shape and angular momentum etc)
obeys precisely the relations that of a Schwarzschild black hole or a
rotating Kerr black hole. Moreover the quantization of these solutions
give rise to a Fermi sea of states whose microstates counting
reproduces precisely the Bekenstein-Hawking entropy of these black
holes.  It was also shown that these microstates distribute uniformly
on the surface of the fuzzy sphere (whose classical large $N$ limit
gives the horizon of black hole) with a density of 2 states per unit
Planck area. Quite amazingly, this is the holographic parton picture
of black hole proposed by Susskind \cite{Susskind:1994vu} in order to
explain the area dependence of the Bekenstein-Hawking entropy. In our
model, this picture is obtained automatically from the fundamental
quantum mechanics. That the holographic interpretation of the
Bekenstein-Hawking entropy can be derived suggested that one may be
able to derive the holographic property of quantum gravity from this
quantum mechanical proposal of quantum gravity.

Let us comment on the connection of our work with the island proposal.
We recall that in the island proposal,   an island appears
and replace a part of the black hole interior
as the number of entangled particle pairs grows. As a result, those particles
which are situated inside the island are no longer part of the black hole
and hence
their entanglement should not  be counted 
toward the entropy. Thus the occurrence of island 
leads to a removal of entanglement and hence
the page curve of the entanglement entropy of 
Hawking radiation.
For us,  the role of island is played the tunneling effect in our model.
Recall that the entanglement entropy is counted by the number $n_2$ of
entangled Bell pairs that connects the thin shell and the outside. Due
to tunneling, an interior particles can tunnel outside, resulting in
an entangled pair that sit entirely outside the black hole and hence
should no longer counted toward the entropy.  As a result, tunneling
leads to a removal of entanglement from the Hawking radiation.  Our
model is based on the assumption that the quantum black hole is given
by a thin shell,
which is justified more recently by the 
construction of fuzzy spheres in  \cite{Chu:2024qil, Chu:2024edh}. 
This seems also to be the case in the island proposal since after
excluding the island from the inner region of black hole, there is
indeed a thin region of space, the quantum extremal surface, that is
left for the black hole. It is suggestive to identify this thin region
of space with the thin shell (or the fuzzy sphere) of quantum space in
our model.  This discussion suggests that two pictures are
complementary descriptions of the same phenomena.  In fact, let us
note that in our model, spacetime is described by a novel quantum
mechanics while entanglement is described in the usual manner. On the
contrary, in the island picture, spacetime is described in terms of
the usual classical geometry while entanglement is captured by a novel
holographic generalized entropy formula. It is interesting to
understand more precisely how island wormhole and tunneling of
entanglement are related, and also how quantum spacetime and
holography are related. The ER=EPR \cite{Maldacena:2013xja} may
provide a key.

We comment that 
the fuzzy solutions of different radii of the theory represent
different minima and are separated from each other by an energy
barrier. Although they are classically stable, there can however be
quantum mechanical tunneling at finite $N$. It is interesting to
analysis if this tunneling process could provide a fully quantum
mechanical description of the origin of the Hawking radiation and the
black hole thermodynamics.
 
In the full quantum gravity description, it is expected that black
hole singularity must be resolved and the black hole S-matrix must be
unitarity. In the semiclassical description, Horowitz and Maldacena
\cite{Horowitz:2003he} have made the interesting proposal that the
unitarity of black-hole evaporation can be reconciled with Hawking’s
semiclassical arguments if a final-state boundary condition is imposed
at the spacelike singularity inside the black hole.  Dissipation,
being a generic feature in a quantum system with an infinite number of
degrees of freedom, has been suggested \cite{Ho:2021huh} as a possible
mechanism for the final state projection.  As we conjecture that
Hawking radiation is described by a tunneling process in the large $N$
quantum mechanics \cite{Chu:2024qil}, it will be interesting to
understand how the boundary state condition arises in the
semi-classical limit there.

{\it Acknowledgments:} We thank Pei-Ming Ho for useful discussion.
C.S.C acknowledge support of this work by NCTS and
the grant 110-2112-M-007-015-MY3 of the National 
Science and Technology Council of Taiwan. 
R.X.M thank support by NSFC grant (No. 11905297).

\appendix

\section{Appendix: Semiclassical analysis}

In a semiclassical estimation,
Hawking radiation reduces the mass of the black hole as
\cite{Page:1976df}
\be \label{M-rate}
\frac{d M}{dt} = - \frac{Q}{3 M^2},
\ee
where $Q=3 \a/G^2$ and $\a$ is some
numerical constant. This gives
\be\label{M3}
M(t) = M_0 \left[ 1- \frac{Q t}{ M_0^3} 
\right]^{1/3}
\ee
and a finite 
lifetime of the black hole
$ t_E = M_0^3/Q$.
In our model, this should come out from the consistent
set of equations
\eq{n1}-\eq{n3} and \eq{Mt}.
Although the complete analysis is quite
complicated, we can see that our model does roughly give
\eq{M-rate} near the final stage of evaporation and so $t_E$
is finite.
In fact, using \eq{n1} and \eq{n2}, we have
  \be\label{Mt-rate}
  \frac{d M}{dt} = \int_{D} - \G_1 \o_1 (n_1+n_2)-  \G_2 \o_2 (n_1+ \nt_2),
  \ee
  where we have ignored the much smaller terms $-(\d_2 \o_1 +\d_1 \o_2 )A$
  on the RHS 
  since $\d \ll \G$ as $\d$ involves an additional vacuum creation rate.  
Near the final stage of evaporation, it is
$\int n_1 \sim \int n_2 \sim O(1)$ and we can use the mean value theorem of
calculus to estimate that
$\int  \G_1 \o_1 n_1= \overline{ \; \G_1 \o_1}  \, \int n_1
\sim 1/M^2$,
where we have used $\o_1 \sim M$ and that $\G \sim \G_0$ for small $M$. Here
the prefactor of the tunneling
rate has a dependence $\G_0 \sim 1/M^3$ coming from the phase space volume. As
a result, \eq{Mt-rate} is consistent with the semiclassical result \eq{M-rate}
at time close to the end point
and so $t_E$ is finite.

To get a better feeling of the time evolution  of the Bell pairs,
let us consider the
approximate mass function $M(t)$ \eq{M3} with $M_0=1,Q=0.05$ so
that $t_E= 20$. In Figure 3,
\begin{figure}[ht]
  \centering
\includegraphics[width=7.5cm]{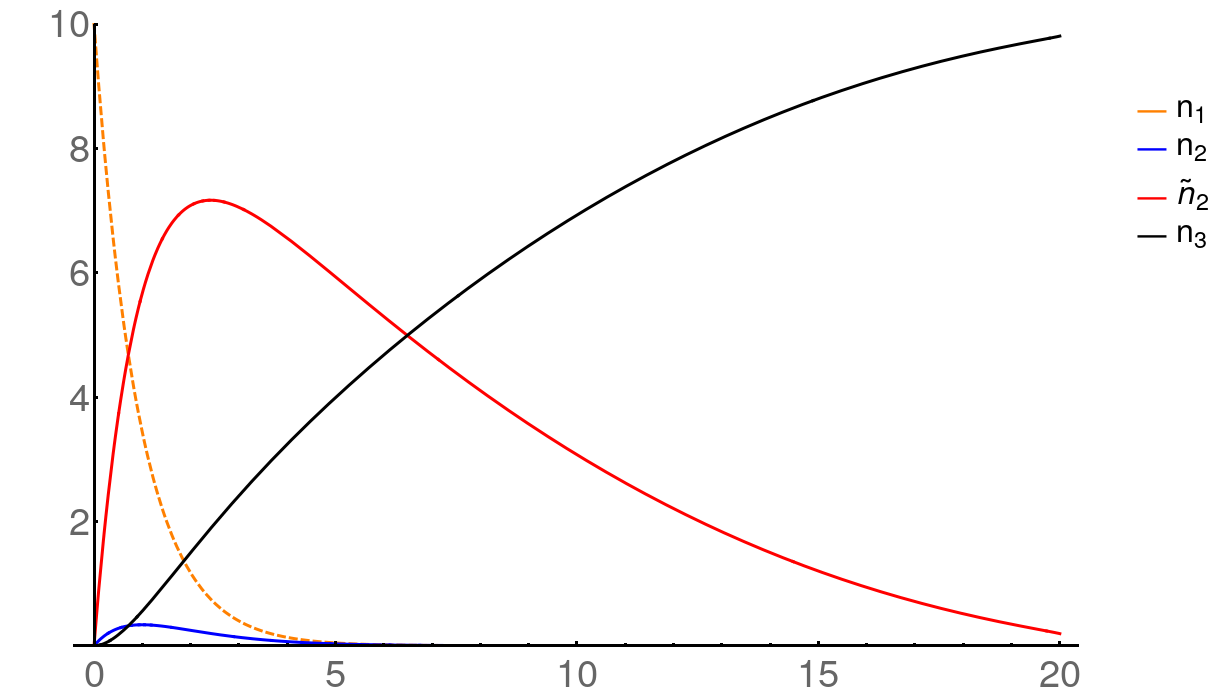}\\
\includegraphics[width=7.5cm]{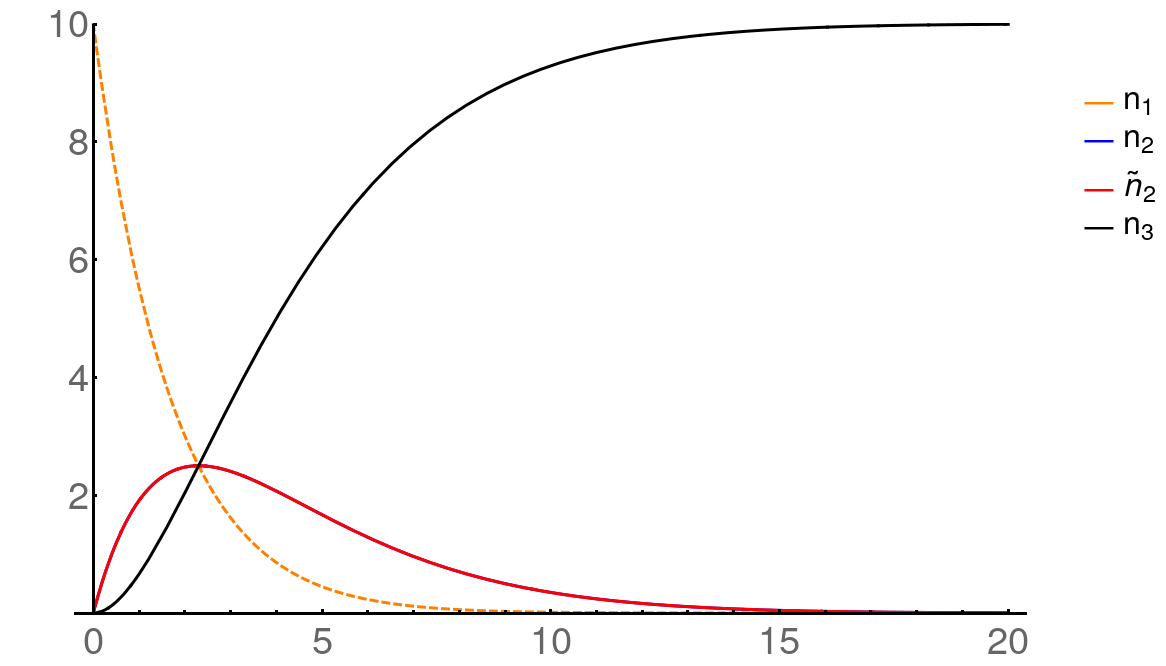}
\caption{Evolution of Bell states for the
  energies $(\o_1,\o_2) = $
  (a) (0.001, 0.1),
    (b) (0.05, 0.05). In case (b), the curves for
    $n_2$ and $\nt_2$ overlap and is represented by the red line.
    The mass \eq{M3} with $M_0=1,Q=0.05$ is adopted for these plots.  }
\end{figure}
we show the plots of $n_1, n_2, \nt_2, n_3$
for two typical set of energies  $(\o_1,\o_2)$ given by
(a): (0.001, 0.1) and
(b): (0.05, 0.05).
Since $\d_a \ll \G_a$, we have taken $\d_a =0$  here for simplicity.
Including $\d_a\neq 0$
won't affect these plots much.
We see that the generic behaviors of the $n_i$'s
discussed above do get captured
very well here, showing that \eq{M3} is a
decent approximation.
However, we can see that for the set (a) of energies,
the time $t_2$ for $n_2$ and $\nt_2$ to decrease to zero is actually slightly
larger than $t_E=20$, showing that
\eq{M3} is not entirely consistent with
\eq{n1}-\eq{n3}.
Nevertheless \eq{M3} is a pretty good approximation and a perturbative
scheme can in principle
be devised to solve the system \eq{n1}-\eq{n3},
\eq{Mt}.


\begin{thebibliography}{00}


\bibitem{Parikh:1999mf}
M.~K.~Parikh and F.~Wilczek,
``Hawking radiation as tunneling,''
Phys. Rev. Lett. \textbf{85} (2000), 5042-5045.


\bibitem{Hawking:1975vcx}
S.~W.~Hawking,
``Particle Creation by Black Holes,''
Commun. Math. Phys. \textbf{43} (1975), 199-220
[erratum: Commun. Math. Phys. \textbf{46} (1976), 206]

\bibitem{Strominger:1996sh}
A.~Strominger and C.~Vafa,
``Microscopic origin of the Bekenstein-Hawking entropy,''
Phys. Lett. B \textbf{379} (1996), 99-104

\bibitem{Hawking:1976ra}
S.~W.~Hawking,
``Breakdown of Predictability in Gravitational Collapse,''
Phys. Rev. D \textbf{14} (1976), 2460-2473

\bibitem{Page:1993wv}
D.~N.~Page,
``Information in black hole radiation,''
Phys. Rev. Lett. \textbf{71} (1993), 3743-3746

\bibitem{Page:2013dx}
D.~N.~Page,
``Time Dependence of Hawking Radiation Entropy,''
JCAP \textbf{09} (2013), 028


   
\bibitem{Engelhardt:2014gca}
N.~Engelhardt and A.~C.~Wall,
``Quantum Extremal Surfaces: Holographic Entanglement Entropy beyond the Classical Regime,''
JHEP \textbf{01} (2015), 073

\bibitem{Ryu:2006bv}
S.~Ryu and T.~Takayanagi,
``Holographic derivation of entanglement entropy from AdS/CFT,''
Phys. Rev. Lett. \textbf{96} (2006), 181602

\bibitem{Hubeny:2007xt}
V.~E.~Hubeny, M.~Rangamani and T.~Takayanagi,
``A Covariant holographic entanglement entropy proposal,''
JHEP \textbf{07} (2007), 062

\bibitem{Penington:2019npb}
G.~Penington,
 ``Entanglement Wedge Reconstruction and the Information Paradox,''
JHEP \textbf{09} (2020), 002

\bibitem{Almheiri:2019psf}
A.~Almheiri, N.~Engelhardt, D.~Marolf and H.~Maxfield,
``The entropy of bulk quantum fields and the entanglement wedge of an evaporating black hole,''
JHEP \textbf{12} (2019), 063


\bibitem{Almheiri:2019hni}
A.~Almheiri, R.~Mahajan, J.~Maldacena and Y.~Zhao,
``The Page curve of Hawking radiation from semiclassical geometry,''
JHEP \textbf{03} (2020), 149

\bibitem{Almheiri:2019qdq}
A.~Almheiri, T.~Hartman, J.~Maldacena, E.~Shaghoulian and A.~Tajdini,
``Replica Wormholes and the Entropy of Hawking Radiation,''
JHEP \textbf{05} (2020), 013

\bibitem{Engelhardt:2017aux}
N.~Engelhardt and A.~C.~Wall,
``Decoding the Apparent Horizon: Coarse-Grained Holographic Entropy,''
Phys. Rev. Lett. \textbf{121} (2018) no.21, 211301


\bibitem{Almheiri:2020cfm}
A.~Almheiri, T.~Hartman, J.~Maldacena, E.~Shaghoulian and A.~Tajdini,
``The entropy of Hawking radiation,''
Rev. Mod. Phys. \textbf{93} (2021) no.3, 035002

\bibitem{Penrose:1964wq}
R.~Penrose,
``Gravitational collapse and space-time singularities,''
Phys. Rev. Lett. \textbf{14} (1965), 57-59
doi:10.1103/PhysRevLett.14.57

\bibitem{tHooft:1993dmi}
G.~'t Hooft,
``Dimensional reduction in quantum gravity,''
Conf. Proc. C \textbf{930308} (1993), 284-296

\bibitem{Susskind:1994vu}
L.~Susskind,
``The World as a hologram,''
J. Math. Phys. \textbf{36} (1995), 6377-6396


\bibitem{Verlinde:2010hp}
E.~P.~Verlinde,
``On the Origin of Gravity and the Laws of Newton,''
JHEP \textbf{04} (2011), 029

\bibitem{Giddings:2006sj}
S.~B.~Giddings,
``Black hole information, unitarity, and nonlocality,''
Phys. Rev. D \textbf{74} (2006), 106005

\bibitem{Kawai:2013mda}
H.~Kawai, Y.~Matsuo and Y.~Yokokura,
``A Self-consistent Model of the Black Hole Evaporation,''
Int. J. Mod. Phys. A \textbf{28} (2013), 1350050 
[arXiv:1302.4733 [hep-th]].

\bibitem{Kawai:2015uya}
H.~Kawai and Y.~Yokokura,
``Interior of Black Holes and Information Recovery,''
Phys. Rev. D \textbf{93} (2016) no.4, 044011 
[arXiv:1509.08472 [hep-th]].

\bibitem{VanRaamsdonk:2013sza}
M.~Van Raamsdonk,
``Evaporating Firewalls,''
JHEP \textbf{11} (2014), 038
[arXiv:1307.1796 [hep-th]].

\bibitem{Akhmedov:2015xwa}
E.~T.~Akhmedov, H.~Godazgar and F.~K.~Popov,
``Hawking radiation and secularly growing loop corrections,''
Phys. Rev. D \textbf{93} (2016) no.2, 024029
[arXiv:1508.07500 [hep-th]].

\bibitem{Ho:2015vga}
P.~M.~Ho,
``The Absence of Horizon in Black-Hole Formation,''
Nucl. Phys. B \textbf{909} (2016), 394-417
[arXiv:1510.07157 [hep-th]].

\bibitem{Good:2016atu}
M.~R.~R.~Good, K.~Yelshibekov and Y.~C.~Ong,
``On Horizonless Temperature with an Accelerating Mirror,''
JHEP \textbf{03} (2017), 013 
[arXiv:1611.00809 [gr-qc]].

\bibitem{Baccetti:2016lsb}
V.~Baccetti, R.~B.~Mann and D.~R.~Terno,
``Role of evaporation in gravitational collapse,''
Class. Quant. Grav. \textbf{35} (2018) no.18, 185005
[arXiv:1610.07839 [gr-qc]].

\bibitem{Mann:2021mnc}
R.~B.~Mann, S.~Murk and D.~R.~Terno,
``Black holes and their horizons in semiclassical and modified theories of gravity,''
Int. J. Mod. Phys. D \textbf{31} (2022) no.09, 2230015
[arXiv:2112.06515 [gr-qc]].

\bibitem{Ho:2022gpg}
P.~M.~Ho and H.~Kawai,
``UV and IR effects on Hawking radiation,''
JHEP \textbf{03} (2023), 002 
[arXiv:2207.07122 [hep-th]].


\bibitem{Page:1976df}
D.~N.~Page,
``Particle Emission Rates from a Black Hole: Massless Particles from an Uncharged, Nonrotating Hole,''
Phys. Rev. D \textbf{13} (1976), 198-206

\bibitem{Almheiri:2012rt}
A.~Almheiri, D.~Marolf, J.~Polchinski and J.~Sully,
``Black Holes: Complementarity or Firewalls?,''
JHEP \textbf{02} (2013), 062

\bibitem{Aschieri:2005zs}
P.~Aschieri, M.~Dimitrijevic, F.~Meyer and J.~Wess,
``Noncommutative geometry and gravity,''
Class. Quant. Grav. \textbf{23} (2006), 1883-1912
 
 
\bibitem{Chu:2023mqi}
C.~S.~Chu and R.~X.~Miao,
``Fermi model of a quantum black hole,''
Phys. Rev. D \textbf{110} (2024) no.4, 046001
doi:10.1103/PhysRevD.110.046001
[arXiv:2307.06164 [hep-th]]. 

\bibitem{Chu:2024qil}
C.~S.~Chu,
``A Matrix Model Proposal for Quantum Gravity and the Quantum Mechanics of Black Holes,'' [arXiv:2406.01466 [hep-th]].

\bibitem{Chu:2024edh}
C.~S.~Chu,
``Quantum Kerr Black Hole from Matrix Theory of Quantum Gravity,''
[arXiv:2406.12704 [hep-th]].

\bibitem{Maldacena:2013xja}
J.~Maldacena and L.~Susskind,
``Cool horizons for entangled black holes,''
Fortsch. Phys. \textbf{61} (2013), 781-811
doi:10.1002/prop.201300020
[arXiv:1306.0533 [hep-th]].

\bibitem{Horowitz:2003he}
G.~T.~Horowitz and J.~M.~Maldacena,
``The Black hole final state,''
JHEP \textbf{02} (2004), 008
[arXiv:hep-th/0310281 [hep-th]].

\bibitem{Ho:2021huh}
P.~M.~Ho,
``Final-State Condition and Dissipative Quantum Mechanics,''
Entropy \textbf{24} (2022) no.10, 1411
doi:10.3390/e24101411
[arXiv:2103.04732 [hep-th]].              

\end{thebibliography}
\end{document}